\documentclass[aps,prd,10pt,twocolumn,preprintnumbers,superscriptaddress,showpacs,floatfix,nofootinbib]{revtex4-1}
\usepackage{textcomp,gensymb}
\usepackage{hyperref}
\usepackage{graphicx}
\usepackage{amsfonts,amsmath,amssymb,bm,bbm}
\usepackage{color,soul}
\usepackage{xcolor}
\usepackage{slashed}
\usepackage[toc,page]{appendix}
\usepackage{flushend}
\usepackage{physics}
\usepackage{graphicx}
\usepackage{dcolumn}
\usepackage{float}

\hypersetup{
    pdfnewwindow=true,      
    colorlinks=true,       
    linkcolor=blue,          
    citecolor=blue,        
    filecolor=blue,      
    urlcolor=blue        
}

\begin{document}
\preprint{
\begin{minipage}{5cm}
\small
\flushright
UCI-HEP-TR-2024-07
\end{minipage}}

\title{Cosmic Ray-Boosted Dark Matter at IceCube}

\author{Christopher V. Cappiello}
\email{cvc1@queensu.ca}
\affiliation{Department of Physics, Engineering Physics, and Astronomy, Queen's University, Kingston, Ontario, K7N 3N6, Canada}
\affiliation{Arthur B. McDonald Canadian Astroparticle Physics Research Institute, Kingston ON K7L 3N6, Canada}
\affiliation{Perimeter Institute for Theoretical Physics, Waterloo, Ontario, N2L 2Y5, Canada}

\author{Qinrui Liu}
\email{qinrui.liu@queensu.ca}
\affiliation{Department of Physics, Engineering Physics, and Astronomy, Queen's University, Kingston, Ontario, K7N 3N6, Canada}
\affiliation{Arthur B. McDonald Canadian Astroparticle Physics Research Institute, Kingston ON K7L 3N6, Canada}
\affiliation{Perimeter Institute for Theoretical Physics, Waterloo, Ontario, N2L 2Y5, Canada}

\author{Gopolang Mohlabeng}
\email{gmohlabe@sfu.ca}
\affiliation{Department of Physics and Astronomy, University of California, Irvine, CA 92697, USA}
\affiliation{Department of Physics, Simon Fraser University, Burnaby, BC, V5A 1S6, Canada}

\author{Aaron C. Vincent}
\email{aaron.vincent@queensu.ca}
\affiliation{Department of Physics, Engineering Physics, and Astronomy, Queen's University, Kingston, Ontario, K7N 3N6, Canada}
\affiliation{Arthur B. McDonald Canadian Astroparticle Physics Research Institute, Kingston ON K7L 3N6, Canada}
\affiliation{Perimeter Institute for Theoretical Physics, Waterloo, Ontario, N2L 2Y5, Canada}

\begin{abstract}
Cosmic ray (CR) upscattering of dark matter is considered as one of the most straightforward mechanisms to accelerate ambient dark matter, making it detectable at high threshold, large volume experiments. In this work, we revisit CR upscattered dark matter signals at the IceCube detector, focusing on lower energy data than was considered before. 
We consider both scattering with electrons and nuclei. In the latter, we include both elastic and deep-inelastic scattering computations. As concrete examples, we consider two benchmark models; Fermion dark matter with vector and scalar mediators. We compare our model projections with the most current constraints and show that the IceCube detector can detect CR-boosted dark matter especially with masses below $\sim$ 100 keV when scattering with electrons and $\sim$ MeV in the nucleon scattering case.
\end{abstract}

\maketitle

\section{Introduction}
\label{sec:intro}
Despite the overwhelming evidence for its existence as well as constituting the majority of the matter in the universe, 
dark matter (DM) is yet to be detected non-gravitationally
~\cite{Bertone:2004pz,Planck:2018vyg,Cooley:2022ufh}. 
This makes its identification one of the most urgent problems in modern science.
For decades, underground direct detection experiments have been searching for Weakly Interacting Massive Particles (WIMPs), that lie in the $\sim$ 1 GeV - 100 TeV mass range.
A possible explanation for it evading detection, is that it could be too light to trigger recoils in typical large detectors that search for WIMPs. This problem can be circumvented by: 1) searching instead for electronic recoils that can be produced by very light DM \cite{Essig:2011nj,Essig:2022dfa}, and 2) accelerating dark matter to high velocity, thus increasing the energy it can deposit in a detector. \\

One such acceleration mechanism is dark matter upscattering by cosmic rays (CRs), proposed in Ref.~\cite{Bringmann:2018cvk}. If dark matter has a nonzero cross section to scatter with nucleons or electrons, cosmic ray-dark matter collisions are an unavoidable consequence. These collisions can accelerate dark matter to relativistic speeds, and for a large enough scattering cross section, the flux of this accelerated dark matter can be detectable. 
Numerous theoretical and experimental works have set limits on (or reported sensitivities to) sub-GeV dark matter based on this process~\cite{Bringmann:2018cvk,Ema:2018bih,Cappiello:2019qsw,Dent:2019krz,Bondarenko:2019vrb,Wang:2019jtk,Dent:2020syp,Ge:2020yuf,Guo:2020drq,Guo:2020oum,Jho:2020sku,Xia:2020apm,Bell:2021xff,Feng:2021hyz,Xia:2021vbz,Alvey:2022pad,Bardhan:2022bdg,Maity:2022exk,Xia:2022tid,PandaX-II:2021kai,CDEX:2022fig,NEWSdm:2023qyb,PROSPECT:2021awi,Super-Kamiokande:2022ncz,PandaX:2024pme}. Several related astrophysical acceleration processes have also been studied in Refs.~\cite{Kouvaris:2015nsa, Hu:2016xas, An:2017ojc, Emken:2017hnp, Calabrese:2021src, Wang:2021jic, Cappiello:2022exa,Alvey:2019zaa, Su:2020zny,Hu:2016xas,Dunsky:2018mqs,Li:2020wyl,Das:2021lcr,Lin:2022dbl}.
Other acceleration mechanisms arising from multi-component dark sectors have also been considered in Refs.~\cite{Agashe:2014yua,Kong:2014mia,Alhazmi:2016qcs,Nam:2020twn,Agashe:2020luo,Kamada:2021muh} and references therein.

In this work, we consider cosmic ray-boosted dark matter interacting in the IceCube detector. IceCube has been considered in this context before, in Refs.~\cite{Guo:2020drq,Guo:2020oum}, the latter of which we follow closely when modeling dark matter-proton scattering. In this article, we build on the work in the current literature in the following ways; We consider dark matter-electron scattering in addition to dark matter-proton scattering, which is yet to be explored at detectors like IceCube. We also include lower energy data than was used in Ref.~\cite{Guo:2020oum} in order to improve sensitivity. 
In addition, we provide limits using two benchmark models: a fermionic dark matter candidate, interacting with the standard model via  either a vector mediator or scalar. We show the limits and prospects of detection at IceCube, in each case.\\

The remainder of this article is largely divided between dark matter-electron and dark matter-nucleon scattering. Sec.~\ref{sec:electron} focuses on dark matter-electron scattering, with four subsections detailing the upscattering by cosmic rays, propagation through the ice to the IceCube detector volume, comparison of the dark matter event rate to IceCube data, and the resulting limits, respectively. Sec.~\ref{sec:proton} begins with a discussion of deep inelastic scattering (DIS), then focuses on dark matter-nucleon scattering using the same framework as the previous Section. In Sec.~\ref{sec:conclusions}, we summarize our results.

\section{Dark Matter-Electron Scattering}\label{sec:electron}

\subsection{Upscattering of Dark matter by Cosmic Ray Electrons}
In this subsection, we consider Galactic, electron cosmic rays colliding with DM particles in the Milky Way halo. We begin with upscattering by electron CRs, but in the following equations we leave the CR species $i$ unspecified, so that they may be used for proton CRs in Sec.~\ref{sec:proton}. The flux of upscattered DM due to collisions with CR species $i$ is written as

\begin{equation}\label{DMdist}
    \frac{\textrm{d}\Phi_{\chi}}{\textrm{d}T_{\chi}}=\int\frac{d\Omega}{4\pi}\int_{l.o.s.}dl\, \frac{\rho_{\chi}(r)}{m_{\chi}} \int dT_i \,\frac{\textrm{d}\sigma_{\chi i}}{\textrm{d}T_{\chi}}\frac{\textrm{d}\Phi_i}{\textrm{d}T_i}\,,
\end{equation}
where $\rho_{\chi}(r)$ is the local DM density, $m_{\chi}$ is the DM mass, and $T_{\chi}$ is the outgoing DM kinetic energy. For the CR spectrum $\frac{\textrm{d}\Phi_i}{\textrm{d}T_i}$, we use the local spectrum of CR electrons measured by VERITAS~\cite{VERITAS:2018iqb}, which extends into the TeV range; because of IceCube's energy range, we are only interested in CRs with at least TeV-scale energies.

If one assumes that the shape of the CR spectrum is independent of the exact position in the Galaxy, then all information about the spatial distribution of DM and CRs can be absorbed into an effective distance parameter $D_{eff}$, resulting in the simplified formula

\begin{equation}\label{DMdist}
    \frac{\textrm{d}\Phi_{\chi}}{\textrm{d}T_{\chi}} = D_{eff}\frac{\rho_{\chi}}{m_{\chi}}\int_{T_i^{min}}^{\infty}dT_i\frac{\textrm{d}\sigma_{\chi i}}{\textrm{d}T_{\chi}}\frac{\textrm{d}\Phi_i}{\textrm{d}T_i}\,.
\end{equation}
In this work, we set $D_{eff}$ = 10 kpc, which is consistent with several recent works on the topic, including several which use GALPROP~\cite{Strong:1998pw} to model the spatial distribution of CRs~\cite{Guo:2020oum,Maity:2022exk,CDEX:2022fig}. This value is consistent with a Navarro-Frenk-White (NFW) profile~\cite{Navarro:1995iw} with a scale radius of 20 kpc. We take $\rho_{\chi} = 0.3$ GeV / cm$^3$ as the local dark matter density.

The cross section $\frac{\textrm{d}\sigma_{\chi i}}{\textrm{d}T_{\chi}}$ depends on the interaction model. In this work, we consider fermionic DM scattering via either a vector or scalar mediator. For a vector mediator, the differential cross section is given by

\begin{equation}
    \frac{\textrm{d}\sigma_{\chi e}}{\textrm{d}T_{\chi}} = \frac{g_{\chi}^2g_e^2m_{\chi}}{4\pi(m_V^2+Q^2)^2}\frac{2(m_e^2+m_{\chi}^2-s)^2 - 2sQ^2 + Q^4}{m_e^4+(m_{\chi}^2-s)^2-2m_e^2m_{\chi}^2-2m_e^2s}\,,
\end{equation}
where $m_V$ is the mediator mass, $s = m_{\chi}^2+m_e^2+2m_{\chi}(T_e+m_e)$ is the center-of-momentum energy squared, and $Q^2 = 2m_{\chi}T_{\chi}$ is the positive four-momentum transfer squared.

For a scalar mediator, the cross section takes the form

\begin{equation}
    \frac{\textrm{d}\sigma_{\chi e}}{\textrm{d}T_{\chi}} =\frac{g_{\chi}^2g_e^2}{32\pi(m_{\phi}^2+Q^2)^2}\frac{(Q^2+4m_{\chi}^2)(Q^2+4m_e^2)}{m_{\chi}(T_e^2+2m_eT_e)}\,.
\end{equation}
When a relativistic electron scatters with a stationary DM particle, the recoil kinetic energy of the DM is given by

\begin{equation}\label{eq:transferredenergy}
    T_{\chi} = \frac{T_e^2 + 2 m_e T_e}{T_e + (m_e + m_{\chi})^2/(2 m_{\chi})}\left( \frac{1 - \cos \theta}{2} \right)\,,
\end{equation}
where $\theta$ is the center-of-momentum frame scattering angle.

Given these cross sections, we can compute the flux of upscattered DM reaching Earth. Because we consider relatively large relativistic cross sections, we assume that any DM propagating toward IceCube from below the horizon is completely blocked, and consider only downgoing events (see Subsection~\ref{subsec:data} for a description of the data). This is a common assumption, used in e.g. Refs.~\cite{PROSPECT:2021awi,Xia:2021vbz} for boosted DM and Ref.~\cite{Emken:2018run} for nonrelativistic DM, and is at most mildly conservative. As IceCube is located at the South Pole, this means that we consider only DM arriving from the southern sky. We conservatively assume that exactly half of the total DM flux arrives from above the horizon. In reality, because the Galactic Center is located in the southern sky and is the direction from which upscattered DM preferentially arrives, more than half the total flux should arrive from above the horizon. However, this choice is conservative, and only affects our limits at the O(10\%) level (see Ref.~\cite{PROSPECT:2021awi}, which considered the daily modulation of such a signal due to the rotation of the Earth).

\subsection{Propagation to IceCube}

The IceCube detector volume is located approximately 1.45 km below the surface of the ice~\cite{IceCube:2014rwe}. This means that we must model the propagation of DM through at least 1.45 km of ice to derive its true energy spectrum at IceCube. This propagation is modeled using a modified version of the nuFATE software~\cite{Vincent:2017svp}. nuFATE was designed for propagation of neutrinos through the Earth, but can be modified to handle relativistic DM propagation by substituting our own differential and total DM cross sections for the neutrino cross sections used in the code. nuFATE assumes that the neutrinos (or in our case, DM particles) travel along straight trajectories, which is a good assumption when the DM kinetic energy $T_{\chi}$ is much greater than the relevant mass scales.

If a DM particle collides with an electron in the outer 90 m of the detector volume, it may be vetoed by the active veto described in Refs.~\cite{IceCube:2013low,IceCube:2014rwe}. To avoid this, we require that a DM particle never scatter in this active veto layer with a deposited energy of over 100 GeV. The effect of this veto on the detected flux is relatively small compared to energy loss in the previous 1.45 km of ice.

Finally, we consider only DM particles that scatter exactly once in the detector volume. At large enough cross sections, particularly when attenuation in the ice overburden is important, DM particles may scatter multiple times, giving rise to a distinctive ``double-bang" signal, or to several interactions that could resemble a track. We ignore all such events, because the data to which we compare consists of all single events. We note, however, that ``double-bang" and track-like signals of new physics at IceCube have been discussed in previous work~\cite{IceCube:2014xnp,IceCube:2015agw,Coloma:2017ppo,Coloma:2019qqj,Mack:2019bps,IceCube:2021eye}.

\subsection{Event Rate and Data Analysis}\label{subsec:data}
We search for signals of CR-boosted DM at IceCube using 641 days of containted neutrino events reported in Ref.~\cite{IceCube:2014rwe}. This reference reports measurements of astrophysical and atmospheric neutrino fluxes down to energies of approximately 1 TeV. This dataset was chosen specifically because of the relatively low energies considered: because of the steepness of the Galactic CR spectra, the flux of upscattered DM also falls with energy. These are contained events, such that the overwhelming downgoing atmospheric muon background is suppressed. Therefore, we chose low-energy IceCube data in order to take advantage of the large IceCube volume while maximizing the detectable flux of DM particles.

Given a DM flux reaching IceCube, we can compute the event rate as

\begin{equation}
    \frac{\textrm{d}R}{\textrm{d}E_{dep}} = \left(\frac{5}{9}\right)N_AT_{exp}\int_{T_{\chi,min}}^{\infty} M_{eff}(E_{dep})\frac{\textrm{d}\sigma_{\chi e}}{\textrm{d}E_{dep}}\frac{\textrm{d}\Phi_{\chi}}{\textrm{d}T_{\chi}}\,,
\end{equation}
where $E_{dep}$ is the energy transferred from the DM to a target electron, which can be obtained by switching $e$ and $\chi$ in Eq.~\ref{eq:transferredenergy}. $T_{exp}$ = 641 days is the exposure time, $N_A$ is Avogadro's Number, the factor of $\frac{5}{9}$ is the number of electrons per nucleon in ice, and $M_{eff}$ is the energy-dependent effective mass. We compute the effective mass from the effective area reported in the supplementary material of Ref.~\cite{IceCube:2014rwe} using the procedure described in Ref.~\cite{Palomares-Ruiz:2015mka}. We compare the spectrum of DM-induced events with the observed events reported in Ref.~\cite{IceCube:2014rwe}. The reported data is divided between the northern and southern sky, with the southern sky defined by $\cos(\theta) > 0.2$, where $\theta$ is the zenith angle. We consider only southern sky events, and consider only DM arriving from within this zenith angle range. As mentioned above, we assume the incoming DM flux is isotropic, a conservative choice given that the Galactic Center is in the southern sky. We perform a binned likelihood analysis, including as background components the expected atmospheric neutrino, astrophysical neutrino, and penetrating muon fluxes reported in Ref.~\cite{IceCube:2014rwe}. Although in Ref.~\cite{IceCube:2014rwe} the power law index $\gamma$ of the astrophysical neutrino flux was allowed to vary, in the energy range relevant for DM signals, the atmospheric neutrino flux inferred by IceCube is always subdominant to other backgrounds. For this reason, we fix $\gamma$ to the best fit value of 2.46 found in Ref.~\cite{IceCube:2014rwe}. We thus fix the shapes of the background rates to be those reported in Ref.~\cite{IceCube:2014rwe}, but allow the normalizations of the astrophysical neutrino, atmospheric neutrino, and penetrating muon fluxes to vary. 
We define the likelihood:

\begin{equation}
    L = \prod_{bins\,i} \frac{e^{-\lambda_i}\lambda_i^{n_i}}{n_i!}\,,
\end{equation}
where $n_i$ is the observed number of events in bin $i$ and $\lambda_i$ is the sum of the expected background and signal. For a given DM mass $m_{\chi}$, and a given value of either $g_{\chi} g_{e}$ or $g_{\chi} g_{q}$, a global maximum likelihood $L_{max}$ is determined by floating the normalizations of all three background components as well as the mediator mass. Then a conditional likelihood $L(m_{med})$ is defined for a given value of the mediator mass $m_{med}$ ($med = \{ Z, \phi \}$ denoting a vector or scalar mediator, respectively). We define the test statistic

\begin{equation}
    -2\Delta \log L = -2(\log L(m_{med}) - \log L_{max})\,,
\end{equation}
and following Ref.~\cite{IceCube:2014rwe}, rule out a value of $m_{med}$ if $-2\Delta \log L > 2.71$.

\begin{figure}
    \centering
    \includegraphics[width=\columnwidth]{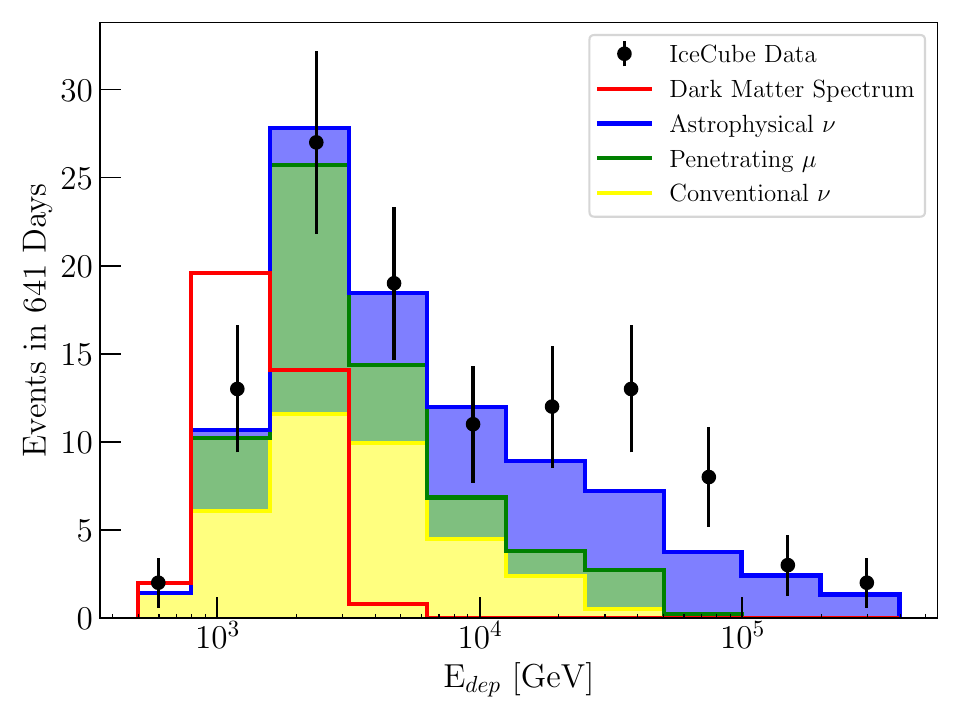}
    \caption{Energy spectrum of southern sky events observed by IceCube up to $\sim$300 TeV (black points), compared to an example spectrum of dark matter events (red). Also shown are the best fit individual neutrino and muon contributions reported by IceCube. This example spectrum is for a vector mediator, with $m_{\chi}$ = 1 keV, $g_{\chi}g_{e}$ = 1, and $M_{Z}$ = 1.16 GeV.
    }
    \label{fig:data}
\end{figure}

In Fig.~\ref{fig:data}, we show the spectrum of southern sky events observed by IceCube in 641 days of data taking, from Ref.~\cite{IceCube:2014rwe}. In comparison, we also show the spectrum of DM events for a vector mediator, with $m_{\chi}$ = 1 keV, $g_{\chi}g_{e}$ = 1, and $M_{V}$ = 1.16 GeV. This set of parameters is close to the limit we set (see the next subsection).

\subsection{Results}

Similar to the approach of Ref.~\cite{Dent:2019krz}, for each value of $m_{\chi}$, we fix the product $g_{\chi}g_e$ and scan over $M_Z$. For the given choices of $m_{\chi}$ and couplings, a value of $M_Z$ can be mapped directly onto a corresponding nonrelativistic cross section, allowing us to set limits on the nonrelativistic DM-electron cross section. The mediator masses at which we set our limits are always much larger than the momentum scale of nonrelativistic scattering, so we define the following nonrelativistic cross sections:

\begin{align}
    &\sigma_{\chi e}^{s} = \frac{g_{\chi}^2g_e^2\mu_{\chi e}^2}{\pi m_{\phi}^4},\\
    &\sigma_{\chi e}^{v} = \frac{4g_{\chi}^2g_e^2\mu_{\chi e}^2}{\pi m_Z^4},\\
    &\sigma_{\chi p}^{s} = \frac{9g_{\chi}^2g_q^2\mu_{\chi p}^2}{\pi m_{\phi}^4},\\
    &\sigma_{\chi p}^{v} = \frac{36g_{\chi}^2g_q^2\mu_{\chi p}^2}{\pi m_Z^4}\,.
\end{align}

\begin{figure}
    \centering
    \includegraphics[width=\columnwidth]{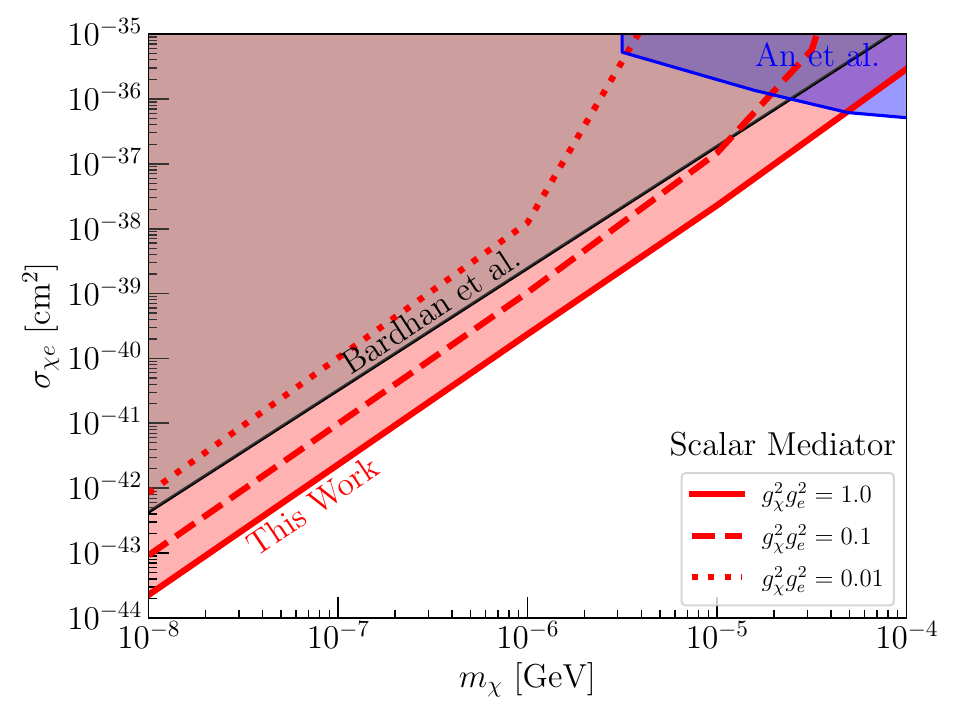}
    \caption{Limits on the nonrelativistic DM-electron cross section for a scalar mediator. Gray is excluded by a search for cosmic ray boosted dark matter at Super-K~\cite{Bardhan:2022bdg}, while blue is excluded based on reflection from the Sun~\cite{An:2017ojc}. The mediator mass at which we set our limits ranges from 30--900 MeV for the three choices of coupling.}
    \label{fig:scalarelectron}
\end{figure}

\begin{figure}
    \centering
    \includegraphics[width=\columnwidth]{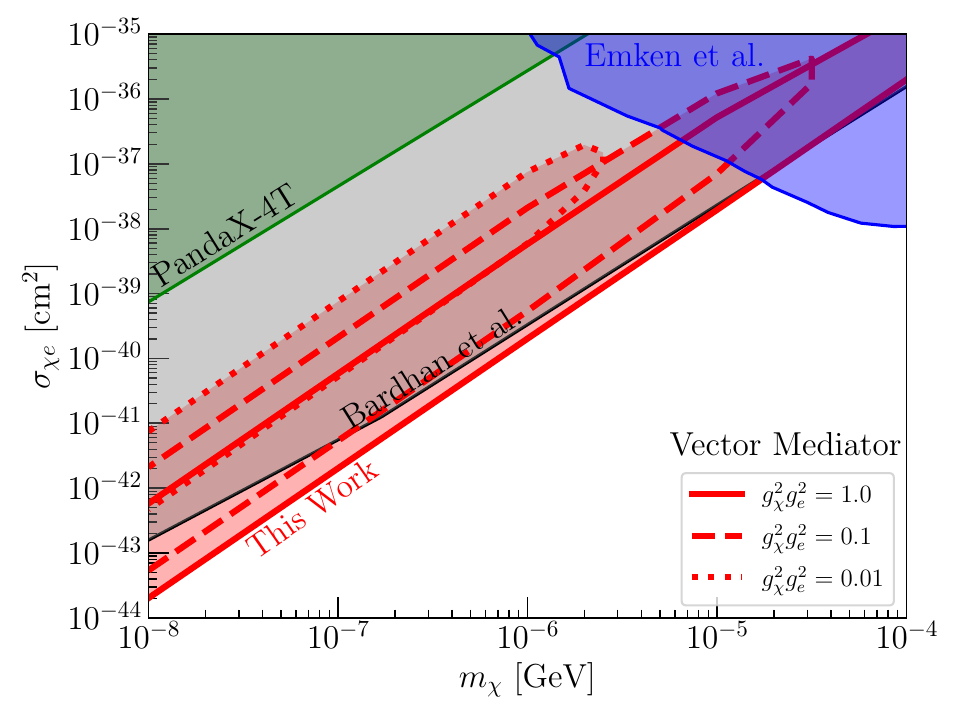}
    \caption{Limits on the nonrelativistic DM-electron cross section for a vector mediator. Gray and green regions are excluded by searches for cosmic ray boosted dark matter at Super-K~\cite{Bardhan:2022bdg} and PandaX-4T~\cite{PandaX:2024pme}, respectively. Blue is excluded based on reflection from the Sun~\cite{Emken:2024nox} (see also Ref.~\cite{An:2021qdl}). The mediator masses constrained here range from 90 MeV--1.25 GeV, across the three choices of coupling.}
    \label{fig:vectorelectron}
\end{figure}

The results for a scalar and a vector mediator are shown in Figs.~\ref{fig:scalarelectron} and~\ref{fig:vectorelectron}, respectively. In both cases, we also show CR upscattering limits from Ref.~\cite{Bardhan:2022bdg}, and solar reflection limits from Refs.~\cite{An:2017ojc} and~\cite{Emken:2024nox} (see also Ref.~\cite{An:2021qdl}). We note that the blazar-boosted DM bounds from Ref.~\cite{Bhowmick:2022zkj} could be stronger than our limits, but they depend on the assumption of unconfirmed dark matter spikes around the blazars in question, and even allowing for that assumption, can vary by orders of magnitude depending on the slope of the spike and the DM annihilation cross section. 

We note that our results for a vector mediator display a ceiling, i.e. a maximum cross section above which IceCube is no longer sensitive. This ceiling is caused by attenuation in the ice above the detector. If the cross section is too large, dark matter will scatter in the ice and lose energy, as described in the previous section. However, our results for a scalar mediator do not display a ceiling---they extend to arbitrarily large cross sections. The reason for this is related to the shape of the two differential cross sections, and to how we vary the cross section. For each of the limits we show, we set the product of the couplings to a constant value, and vary the cross section by changing the mediator mass (as done in Ref.~\cite{Dent:2019krz}). 
But, because the scattering occurs at relativistic energies, it is possible for the momentum transfer to be large compared to the mediator mass. When this is the case, decreasing the mediator mass---which increases the \emph{nonrelativistic} cross section, which we plot in order to compare to direct detection experiments---does not affect the \emph{relativistic} cross section, which determines the scattering rate. When we examine the differential cross sections for the scalar and vector mediator cases, we find that scattering via a scalar mediator is heavily weighted toward large momentum transfer (for a given value of $s$), while scattering via a vector mediator is weighted toward small momentum transfer. 
This means that, as we decrease the mediator mass, the scalar case quickly reaches a point where reducing the mediator mass does not increase the scattering rate and thus attenuation. In the vector case, however, making the mediator mass small does lead to a large scattering rate in the ice above the detector, producing the aforementioned ceiling.

\section{Dark Matter-Nucleon Scattering}\label{sec:proton}

We next turn to the case of DM-proton scattering. the primary complication compared to the electron case is that the energies we consider are well into the regime of deep inelastic scattering. We therefore begin by describing the formalism of both elastic and deep inelastic scattering (DIS).

\subsection{Deep Inelastic Scattering}

In this subsection, we closely follow the work of Ref.~\cite{Guo:2020oum}. We assume that DM couples equally to all quarks, with the couplings to quarks and protons related simply by $g_q = g_p/3$. 

In the low momentum transfer regime, the scattering is elastic. The elastic scattering cross section between DM and a proton can be written as (see equivalent expressions in Refs.~\cite{Dent:2019krz,Cao:2020bwd})

\begin{equation}
    \frac{\textrm{d}\sigma}{\textrm{d}Q^2} = \frac{g_p^2g_{\chi}^2}{4\pi\beta^2}\frac{1}{(Q^2+m_Z^2)^2}\left(1-\beta^2\frac{Q^2}{Q^2_{max}}+\frac{Q^4}{8m_p^2E_{\chi}^2}\right)G_p^2(Q^2)\,.
\end{equation}
Here $\beta$ is the incoming DM velocity, $E_{\chi}$ is the incoming DM energy, and $Q^2 = -q^2$. $G_p(Q^2)$ is the proton form factor, for which we use the form

\begin{equation}
    G_p(Q^2) = \frac{1}{(1+Q^2/\Lambda_p^2)^2}\,,
\end{equation}
with $\Lambda_p$ = 770 MeV~\cite{2004ADNDT..87..185A}. $Q^2_{max}$ is given by

\begin{equation}
    Q^2_{max}=\frac{4(\frac{E_{\chi}^2}{m_{\chi}^2}-1)m_{\chi}^2m_p^2}{m_{\chi}^2+m_p^2+2m_pE_{\chi}}\,.
\end{equation}
This form of the cross section can be shown to be equivalent to the vector mediator case shown for the electron above. However, this factorization more closely matches the form of the DIS cross section given below.

In the high momentum transfer regime, we need the DIS cross section, given by \footnote{Note that a typo exists in Ref.~\cite{Guo:2020oum}; the formula used here is more directly adapted from Ref.~\cite{Formaggio:2012cpf}.}

\begin{equation}
\begin{split}
    \frac{\textrm{d}\sigma}{\textrm{d}\nu\textrm{d}Q^2} = \frac{1}{2\nu}\frac{g_q^2g_{\chi}^2}{4\pi(Q^2+m_Z^2)^2}\,\times\\
    \left(1-\frac{\nu}{E_{\chi}}+\frac{\nu^2}{2E_{\chi}^2}-\frac{Q^2}{4E_{\chi}^2}\right)F_2(x,Q^2)
\end{split}
\end{equation}
Here,

\begin{equation}
    F_2(x,Q^2) = x \sum_i \left[f_i(x,Q^2) + \overline{f}_i(x,Q^2)\right]\,,
\end{equation}
where the sum is over all quark flavors and $f_i (\overline{f}_i)$ denote the parton distribution function (PDF) of a given quark (antiquark) flavor within the nucleon. To model scattering with neutrons, we simply exchange the PDF contributions from the up and down quarks relative to the proton, leaving all other quarks unchanged. $\nu$ is the energy lost by the incoming particle\footnote{In terms of the inelasticity $y$, $\nu = \frac{s - m_p^2 -m_{\chi}^2}{2m_p} \,y.$ } (in the proton's rest frame), and $x$ is the traditional Bjorken $x$. We use the CT10 NLO proton PDFs~\cite{Lai:2010vv} obtained using LHAPDF~\cite{Buckley:2014ana}.

For the scalar case, in the ultrarelativistic limit, the DIS cross section is given by

\begin{equation}
    \frac{\textrm{d}\sigma}{\textrm{d}\nu\textrm{d}Q^2} = \frac{g_q^2g_{\chi}^2}{16\pi}\frac{\nu}{E_{\chi}^2(Q^2+m_Z^2)^2}xf(x)\,.
\end{equation}

In the intermediate regime, resonance excitations are important. However, theoretical modeling of such excitations requires in-depth modeling of the transport of hadronic states within the nucleus (see, for example, Ref.~\cite{Buss:2011mx}). These excitation reactions can be computed for the Standard Model using specialized codes, but mapping these results onto a given DM model is nontrivial (see however Refs.~\cite{Guo:2020oum,Alvey:2022pad}). Neglecting them should have minimal effect on our results: as we will argue below, the upscattering of light DM should be dominated by quasielastic scattering, while the deposition of energy in the IceCube volume will be dominated by DIS.

\subsection{Dark Matter Flux at IceCube}

In this work, we consider upscattering by proton CRs, and no heavier nuclei. Our analysis requires the CR proton spectrum as an input (see Eq.~\ref{DMdist}), which we obtain by interpolating the DAMPE measurement of the proton flux, which extends from 40 GeV to 100 TeV~\cite{DAMPE:2019gys}. When a proton collides with a DM particle, the upscattered flux is once again computed via Eq.~\ref{DMdist}, but with the elastic cross section replaced with the sum of the elastic and inelastic cross sections. 

\begin{figure}
    \centering
    \includegraphics[width=\columnwidth]{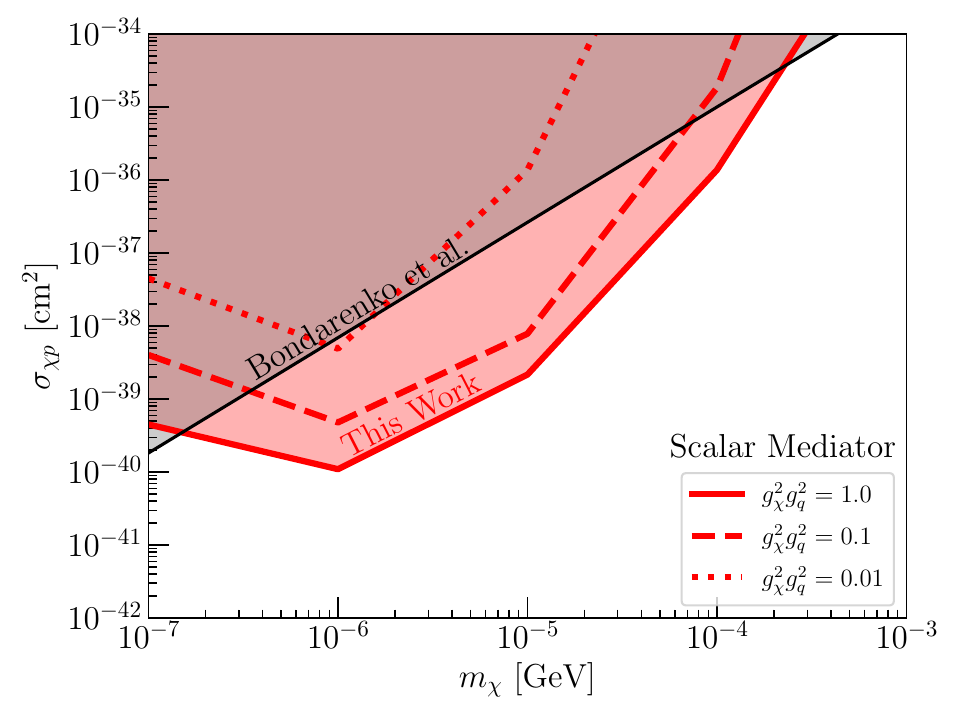}
    \caption{Limits on the nonrelativistic DM-proton cross section for a scalar mediator. The gray exclusion region is based on cosmic ray boosted dark matter interacting in XENON1T~\cite{Bondarenko:2019vrb}. The mediator mass at which we set our limit ranges from 40 MeV to 2.7 GeV, across the three choices of coupling.}
    \label{fig:scalarproton}
\end{figure}

Despite the large energies involved, this upscattering is largely dominated by elastic scattering. Because of the steep slope of the CR spectrum, our limits will be set by the low-energy end of the IceCube data, around 1 TeV (as shown in Fig.~\ref{fig:data}). In order for inelastic processes to start to be important, the momentum transfer must be at least of order $\sim$100 MeV. As we will see below, the limits that we set fall in the mass range $M_{\chi} \lesssim 1$ MeV. In the majority of this mass range, $q^2 = 2m_{\chi}E_{\chi}$ is much less than 100 MeV for kinetic energies of order 1 TeV. This confirms that for most of our parameter space, upscattering is indeed governed by elastic scattering, and neglecting inelastic effects is mildly conservative. 

As above, we propagate the upscattered DM through 1.45 km of ice using the modified nuFATE code, considering only particles arriving from the southern sky. To model attenuation, only DIS is significant. The reason is that, with energies of order 1 TeV, elastic scattering and resonance excitations will only cause energy loss up to order 1 GeV, whereas in DIS a particle can lose approximately all of its energy in a single collision. For this reason, we only include DIS in our attenuation calculation. We again require that a DM particle never deposit over 100 GeV of energy in the veto region surrounding the detector, and exclude any particles that scatter more than once in the remaining IceCube volume.

\subsection{Results}

As with attenuation, we consider only DIS for DM depositing energy within the IceCube volume. We set limits on DM-proton scattering using the same statistical approach as in the electron case, requiring that the DM event rate never be significantly larger than the total observed event rate. 

\begin{figure}
    \centering
    \includegraphics[width=\columnwidth]{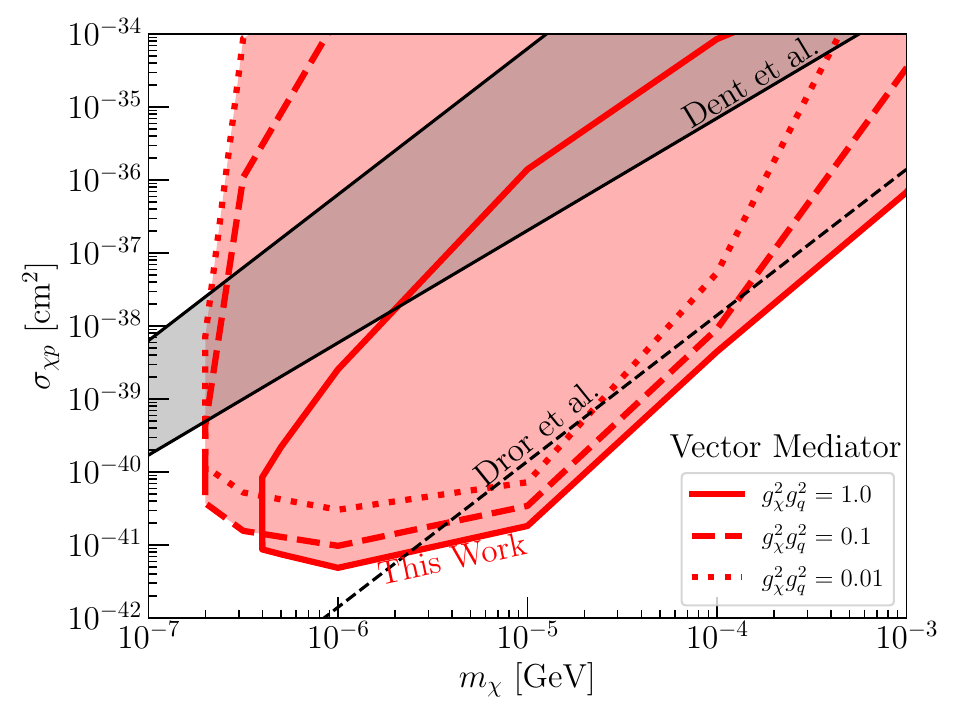}
    \caption{Limits on the nonrelativistic DM-proton cross section for a vector mediator. The gray exclusion region is both based on cosmic ray boosted dark matter interacting in XENON1T~\cite{Dent:2019krz}, while the dashed black line is a limit on new vector mediators from Ref.~\cite{Dror:2017ehi}. The mediator masses we exclude range from 11 MeV--3.9 GeV, across the three choices of coupling.}
    \label{fig:vectorproton}
\end{figure}

Figures~\ref{fig:vectorproton} and~\ref{fig:scalarproton} show our limits on the nonrelativistic DM-proton scattering cross section, for a vector and scalar mediator, respectively. As for electrons, we fix the value of the product of the couplings and vary $m_Z$ in order to vary the cross section. As in the case for electrons, we find a ceiling for the vector case but not the scalar case. 

We compare our results with the exclusion region from Ref.~\cite{Dent:2019krz} and a compilation of exclusion regions from Ref.~\cite{Bondarenko:2019vrb}. Note that in the latter case, the limits are set by fixing the mediator mass and varying the couplings, and we have combined their exclusion regions for a selection of different masses. The exclusion region of Ref.~\cite{Dent:2019krz} follows our approach, with $g_{\chi}g_{p}$ set to 1.

Recently, Ref.~\cite{Bell:2023sdq} pointed out that the relatively large couplings and small mediator masses typically assumed in studies of proton- or nucleus-upscattered DM are subject to various constraints, including limits from meson decay and stellar cooling. In the case of a scalar mediator, the constraint cited from Ref.~\cite{Knapen:2017xzo} is difficult to map onto our results, as the models considered couple dominantly to either the top quark or to gluons, neither of which we consider in our work. However, the constraint on a baryon number-coupled vector mediator from Ref.~\cite{Dror:2017ehi} maps directly onto our parameter space. In Fig.~\ref{fig:vectorproton}, we plot the resulting limit from Ref.~\cite{Dror:2017ehi} for a mediator mass of 1 GeV, taking the more conservative case where the mediator does not decay invisibly. At this mass, the leading limit comes not from meson decay, but from anomaly cancellation considerations~\cite{Dobrescu:2014fca}. We find that our limits are comparable to this constraint, and as they arise from entirely different physics, are highly complementary. The extension of direct detection constraints down to the keV scale for both DM-electron and DM-nucleon scattering is a major goal driving the development of new, low-threshold detectors---see Ref.~\cite{Essig:2022dfa} for a recent review. The limits we set overlap significantly with the far-term goals laid out in that white paper, and push to even lower masses.

\section{Conclusions}\label{sec:conclusions}

In this work we investigated the prospects for detecting cosmic ray-boosted dark matter at the IceCube detector, focusing on relatively low-energy IceCube data. In our analysis, we studied scattering with both electrons and nucleons, including important DIS calculations in the latter. As concrete examples, we applied the data to two benchmark simplified models of fermion dark matter via vector and scalar mediators. In each scenario, we computed the DM flux from cosmic ray upscattering and derived limits on the non-relativistic scattering cross-section. 

In the case of electron scattering, our limits are comparable with, or stronger than, the strongest existing limits on cosmic ray-boosted dark matter, depending on the choice of coupling. For proton scattering, our limits are substantially stronger than existing cosmic ray-boosted dark matter limits over a wide range of masses. And for the case of a vector mediator coupled to baryon number, our limits are comparable to independent limits on the couplings of such mediators based on meson decay and anomaly cancellation considerations.

Throughout this work, we have assumed dark matter arrives isotropically. More detailed modeling of the Galaxy's geometry would improve our limits, but probably not beyond the O(10\%) level. In the energy range from 1--10 TeV, there exists an offset between the cosmic ray proton spectra measured by DAMPE~\cite{DAMPE:2019gys} and ATIC~\cite{2009BRASP..73..564P} and that measured by CREAM~\cite{Yoon:2017qjx}. However, this discrepancy is at most about 20\%, and should not have a large impact on our results.

We expect to obtain improved constraints in the future with the IceCube Upgrade which will optimized for neutrino detection at energies below TeV by deploying 7 new strings near the bottom center of the existing IceCube with denser photosensors~\cite{Ishihara:2019aao}.
Given the steepness of the cosmic ray proton and electron spectra, and of the resulting dark matter spectra, this improvement of detection at the GeV range could increase IceCube's sensitivity to upscattered dark matter. On the other hand, as our limit is set by comparing so roughly the square root of the observed event rate, and the dark matter event rate is proportional to the cross section squared, our limit on the cross section will only scale with the fourth root of the effective area. Therefore a significant improvement to the limit would require a multiple order of magnitude increase in the effective area toward low energies.

\begin{acknowledgments}
We are grateful to Matthias Danninger, Gang Guo, Liangliang Su, Lei Wu, and Meng-Ru Wu for helpful discussions. CVC, QL, and ACV were supported by the Arthur B. McDonald Canadian Astroparticle Physics Research Institute. Research at Perimeter Institute is supported by the Government of Canada through the Department of Innovation, Science, and Economic Development, and by the Province of Ontario. GM acknowledges support from the UC office of the President through the UCI Chancellor's Advanced Postdoctoral Fellowship, the U.S. National Science Foundation under Grant PHY-2210283 and the National Sciences and Engineering Research Council of Canada. ACV acknowledges further support from NSERC, CFI and the Ontario Ministry of Colleges and Universities.
\end{acknowledgments}

\bibliography{main}


\end{document}